\documentclass[a4paper,12pt,doublespace]{article}
\usepackage[utf8]{inputenc}
\usepackage{amsmath}
\usepackage{amssymb}
\usepackage[font={footnotesize}]{caption}
\usepackage{float}
\usepackage{physics}
\usepackage{mathtools}
\usepackage{authblk}
\usepackage{xcolor}
\usepackage{fancyhdr}
\usepackage{graphicx} 
\usepackage{subfigure}
\usepackage{cite}

\title{Evidence of fractal structures in hadrons}

\author[1]{Rafael P. Baptista}
\affil[1]{Instituto de Física - Universidade de São Paulo - Brazil}
\author[1]{Lucas Q. Rocha}

\author[2]{D. P. Menezes}
\affil[2]{Departamento de F\'{\i}sica, CFM-Universidade Federal de Santa Catarina, Florian\'opolis, SC-CP. 476-CEP
88.040-900, Brazil}

\author[3]{Luis A. Trevisan}
\affil[3]{Departamento de Matematica e Estatística - Universidade Estadual de Ponta Grossa, Brazil}

\author[4,5,6]{Constantino Tsallis}
 \affil[4]{Centro Brasileiro de Pesquisas F\'isicas and National Institute of Science and Technology of Complex Systems, Rua Xavier Sigaud 150, Rio de Janeiro-RJ 22290-180, Brazil \hspace{3.5cm} 
 $^{5}$Santa Fe Institute, 1399 Hyde Park Road, Santa Fe, 87501 NM, USA \hspace{10.5cm}
$^{6}$Complexity Science Hub Vienna - Josefst\"adter Strasse 39,  
1080 Vienna, Austria}

\author[1]{Airton Deppman}

\date{July 2023}

\begin{document}

\maketitle

\abstract{
 This study focuses on the presence of (multi)fractal structures in confined hadronic matter through the momentum distributions of mesons produced in proton-proton collisions between 23~GeV and 63~GeV. The analysis demonstrates that the $q$-exponential behaviour of the particle momentum distributions is consistent with fractal characteristics, exhibiting fractal structures in confined hadronic matter with features similar to those observed in the deconfined quark-gluon plasma (QGP) regime. Furthermore,  the systematic analysis of meson
production in hadronic collisions at energies below 1 TeV
 suggests that specific fractal parameters are universal, independently of confinement or deconfinement, while others may be influenced by the quark content of the produced meson. These results pave the way for further research exploring the implications of fractal structures on various physical distributions and offer insights into the nature of the phase transition between confined and deconfined regimes.
}

\section{Introduction}

The investigation of strongly interacting particles has always faced the challenge of dealing with regimes where perturbative QCD (pQCD) cannot provide accurate results. The structure of hadrons and the existence of quark-gluon plasma (QGP) are notorious examples of systems where non-perturbative QCD (npQCD) must be considered.

The large amount of data provided by high-energy colliders in the TeV energy region has revealed several aspects of strong interactions in npQCD. One of the most ubiquitous features is the q-exponential behaviour observed in particle momentum distributions~\cite{T-Biro2, Cleymans-Azmi,Wilk-Wlodarczyk1, Cleymans-Parvan}, suggesting that nonextensive statistical mechanics, based on the nonadditive entropy S$_q$,~\cite{Tsallis, TsallisBookV2} is the appropriate framework to study the complex system formed during collisions. The emergence of non-extensive statistics (or {\it $q$-statistics} for short) in Yang-Mills theories has been demonstrated in ~\cite{DMM-PRD-2020}, which uses self-energy interactions and the renormalization group equation to show how thermofractal structures can emerge in the interacting fields. These thermofractal structures give rise to non-extensive statistics and $q$-exponential distributions~\cite{Deppman-PRD-2016}.

Assuming that the statistical mechanics grounded on S$_q$ is the correct framework to study the thermodynamics of QCD systems, it is natural to extend it to describe high-energy processes as well. Hagedorn's Self-Consistent Thermodynamics~\cite{Hagedorn1} was generalized by incorporating non-extensive statistics~\cite{Deppman-2012}, resulting in a critical temperature, similar to the original theory, but with a new formula for the hadron mass spectrum. The new formula provides a better description compared to the original Hagedorn's formula, accurately describing the mass spectrum from the heaviest known hadrons down to the pion mass~\cite{Marques-Andrade-Deppman-2013}. This result suggests the existence of fractal structures in hadrons beyond those already investigated in the deconfined regime. Could the confined quark matter inside hadrons show fractal aspects? The present work offers a systematic analysis of meson production in hadronic collisions at energies below 1 TeV to address this question. It will be shown that the momentum distributions of produced mesons resulting from collisions in the range of 23 GeV to 63 GeV exhibit many features expected in thermofractals.

\begin{figure}[t]
\centering	
 \centering\begin{subfigure}{}
     \includegraphics[width=0.5\linewidth]{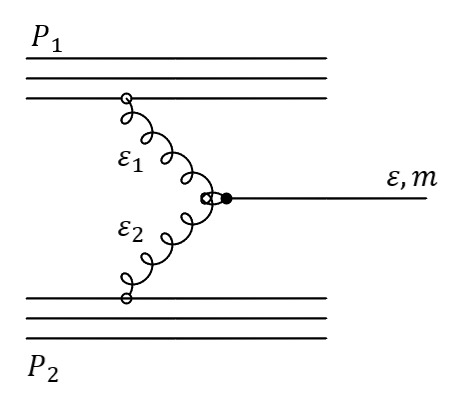} 
 \end{subfigure}	
 \caption{Example of the simplest subprocess for meson production in $pp$ collision.}
	\label{fig:Subprocess}
\end{figure}

A subproduct of this work is the reconciliation of QCD with the early theoretical approaches proposed in the 1960s by Chew, Frautshi~\cite{Chew-Frautschi1} and Hagedorn, among others. The self-consistent calculation, also known as the bootstrap approach, for describing hadrons was an important line of research whose prominence diminished with the emphasis given to
quarks and QCD. The understanding of fractal structures in Yang-Mills Theories unifies the two approaches, clarifying and simplifying several aspects of Quantum Chromodynamics.

The results obtained in the present study unveil new possibilities for investigating the effects of fractal structures in hadronic processes. Experimental data on the hadron structure are produced in HERA, JLab, and NICA, among others, while the forthcoming Electron-Ion Colliders (EIC) offer promising prospects for experimentally probing the intricate nature of the hadronic structure~\cite{Bruning2022-kd, Abelleira_Fernandez2012-xl}. Consequently, the present work appears to hold substantial relevance for the advancement of these colliders. Moreover, given the presence of fractal structures in the Quark-Gluon Plasma (QGP), the identification of similar structures in the hadronic phase may impose constraints on the confinement-deconfinement phase transition. Thus, the present findings should  contribute significantly to the understanding of fundamental characteristics associated with hadronic matter.

The non-extensive statistics has already been used in hadronic gas models~\cite{Roynek2016, Cardoso2017, Andrade2020}. The present approach differs from the previous works in the use of the running-coupling and microscopic calculation of the subprocess involved in hadron collisions.

\section{Methods}

For the investigation of the thermofractal structures in hadrons, this work analyses the momentum distributions of mesons produced in $pp$ collisions. The data is collected in the centre-of-mass frame at approximately zero rapidity, where the colliding protons have momentum $P$ and total energy $E=\sqrt{s}$~\cite{Alper1975}. In the relevant subprocess, two quarks interact and form a new meson, while the other quarks are considered spectators. It is assumed that almost all energy of the process is consumed in the production of the meson, so the momentum of the remaining system is negligible. This assumption limits, to some extent, the power of the present analysis, but it does not interfere with the pursued objectives. In this configuration, if $\varepsilon$ is the produced meson energy and $\varepsilon_1$,  $\varepsilon_2$ are the energies of the interacting quarks, then  $\varepsilon_1=\varepsilon_2=\varepsilon/2$.

Fig.~\ref{fig:Subprocess} shows an example of the meson production process to assist the discussion of some relevant aspects that will affect the analysis of the experimental data. Since we are in the confined regime of the quark matter, the meson production involves at least three vertices. Using the results of the thermofractal theory for QCD, we have that each vertex involves a running coupling given by
\begin{equation}
    g(\varepsilon)= G_o e_q(\varepsilon_1,\lambda_1;q)e_q(\varepsilon_2,\lambda_2;q) \,,
\end{equation}
where $G_o$ is a constant associated with the maximum value of the coupling, the $q$-exponential distribution is given by
\begin{equation}
    e_q(\varepsilon,\lambda;q) = \left[1+(q-1)\frac{\varepsilon}{\lambda}\right]^{\frac{-1}{q-1}} \,,
\end{equation}
with $\varepsilon=\varepsilon_1+\varepsilon_2$, where $\varepsilon_1$ and $\varepsilon_2$ are the energy carried by the interaction partons in each proton, represented by the indexes $1$ and $2$, $\lambda$ is a scale parameter (effective temperature) and $q$ is the entropic index associated with S$_q$.

For fractal structures, one expects to observe the characteristic $q$-exponential distributions due to the form of the running coupling. Besides the form of the distribution, it is expected that~\cite{DMM-PRD-2020}
\begin{equation}
    \frac{1}{q-1}=\frac{11}{3}N_c-\frac{4}{3}\frac{N_f}{2} \,,
\end{equation}
where $N_c$ and $N_f$ are, respectively, the numbers of colours and flavours. With $N_c=3$ and $N_f=6$, the formula above gives $q=8/7 \simeq 1.14$, in agreement with high-energy experimental data analyses.

The work performs three scenarios of systematic analyses, as discussed below, where the experimental data is fitted by three different models, with increasing physical significance, where the momentum distribution is given by: 
\begin{enumerate}
    \item  A single $q$-exponential function, with
    \begin{equation}
        \frac{d^3 \sigma}{dp^3}= \sigma_o e_q(\varepsilon,\lambda;q) \,, \label{eq:mod1}
    \end{equation}
    where $q$, $\sigma_0$, and $\lambda$ are adjustable parameters.
    \item  The product of three $q$-exponential functions with
     \begin{equation}
        \frac{d^3 \sigma}{dp^3}= \sigma_o e_q(\varepsilon_1,\Lambda;\bar{\bar{q}}) e_q(\varepsilon_2,\Lambda;\bar{\bar{q}}) e_q(\varepsilon,\lambda;q) \,, \label{eq:mod2}
    \end{equation}
    where $\varepsilon_1=\varepsilon_2=\varepsilon/2$, $(\bar{\bar{q}}-1)^{-1}=[(q-1)/q]^{-1}$, and the parameters $\sigma_o$, $\lambda$ and $\Lambda$ are adjusted. The parameter $\bar{\bar{q}}$ takes into account that the parton momentum distributions in the hadron emerge in a complex interaction of an equilibrated system~\cite{Megias2015}. The use of $\bar{\bar{q}}$ can be controversial. A thorough discussion about this point can be found in Refs.~\cite{Cleymans-Parvan, Wilk-Wlodarczyk1, T-Biro2, Rocha2022}, but it can be advanced that results very similar to those shown here can be obtained by using $\bar{\bar{q}}=q$.

    \item The product of three $q$-exponential functions, as in Scenario 2, but where $\lambda$ is fixed at a value that will be found in Scenario 2 and the parameters $\sigma_o$ and $\Lambda$ are adjusted;
    
\end{enumerate}

\begin{figure}[ht]

\centering
\begin{subfigure}{}
\includegraphics[width=0.3\linewidth]{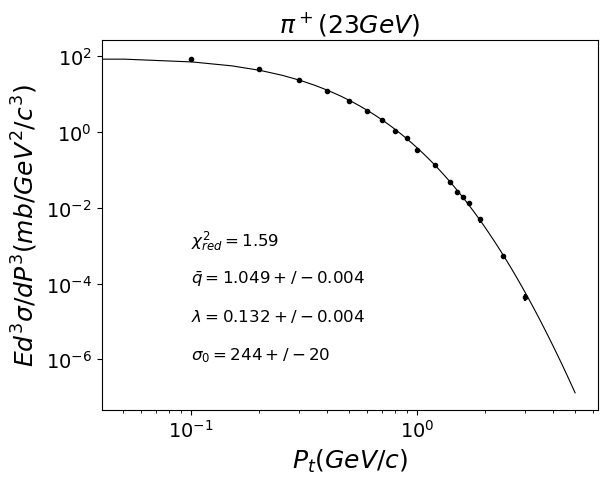} 
 \end{subfigure}
\centering
\begin{subfigure}{}
\includegraphics[width=0.3\linewidth]{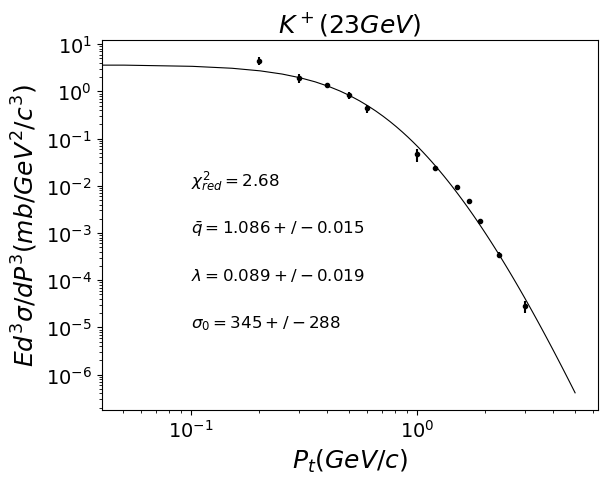} 
 \end{subfigure}
\centering
\begin{subfigure}{}
\includegraphics[width=0.3\linewidth]{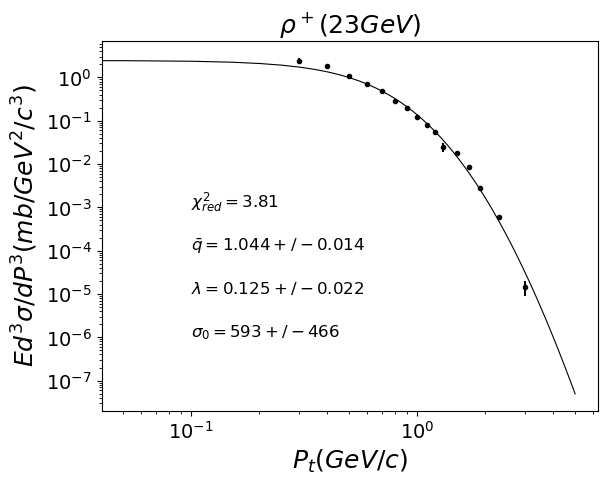} 
 \end{subfigure}
 \centering\begin{subfigure}{}
    \includegraphics[width=0.3\linewidth]{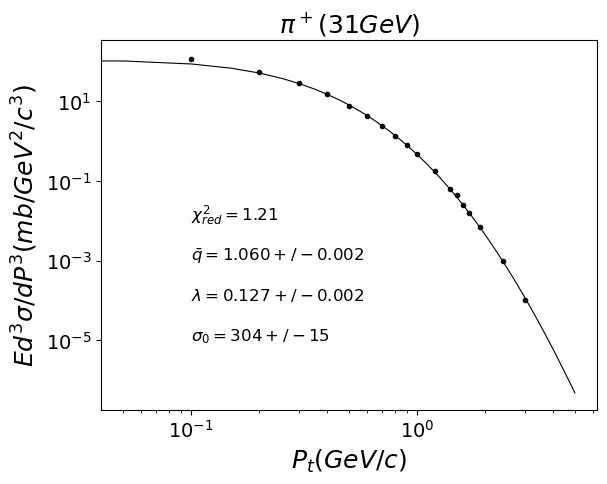}
 \end{subfigure}	
  \centering\begin{subfigure}{}
    \includegraphics[width=0.3\linewidth]{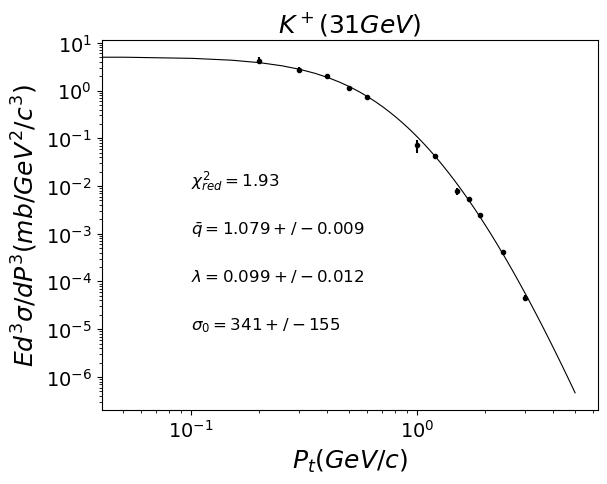}
 \end{subfigure}	
 \centering\begin{subfigure}{}
    \includegraphics[width=0.3\linewidth]{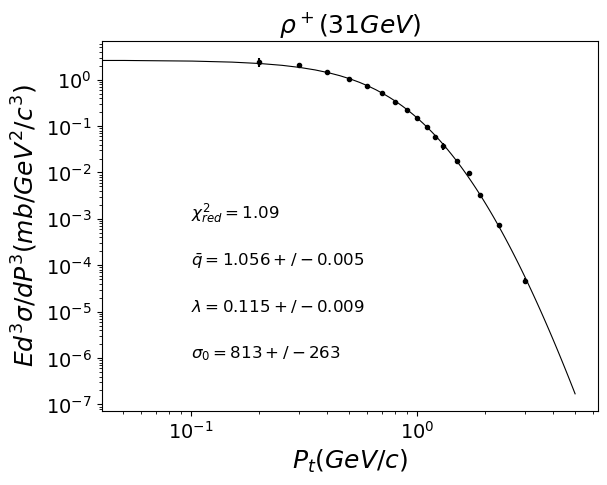}
 \end{subfigure}	
 \centering\begin{subfigure}{}
    \includegraphics[width=0.3\linewidth]{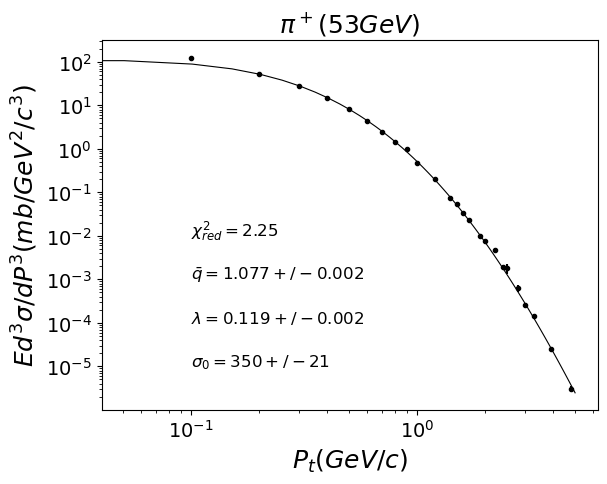} 
 \end{subfigure}	
  \centering\begin{subfigure}{}
    \includegraphics[width=0.3\linewidth]{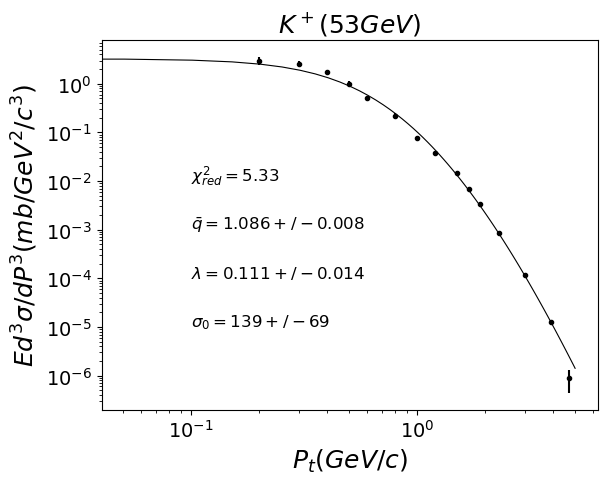} 
 \end{subfigure}	
  \centering\begin{subfigure}{}
    \includegraphics[width=0.3\linewidth]{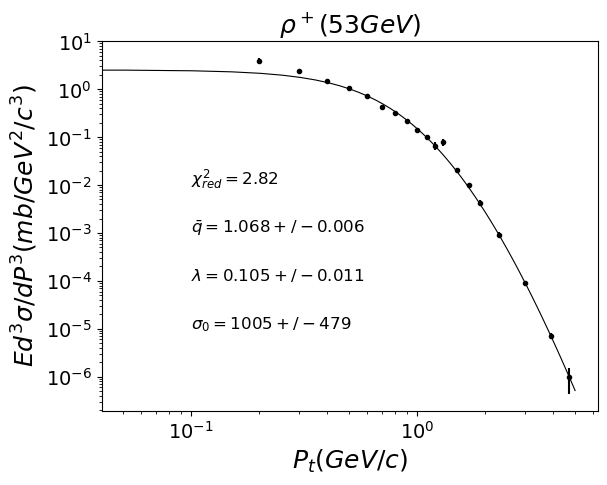} 
 \end{subfigure}	
 \centering\begin{subfigure}{}
     \includegraphics[width=0.3\linewidth]{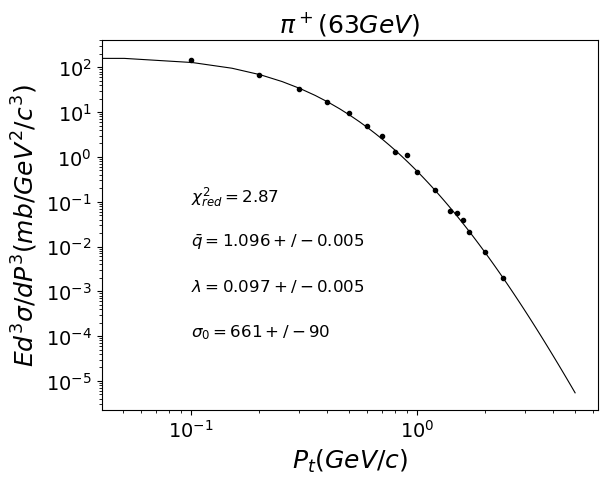} 
 \end{subfigure}	
 \centering\begin{subfigure}{}
     \includegraphics[width=0.3\linewidth]{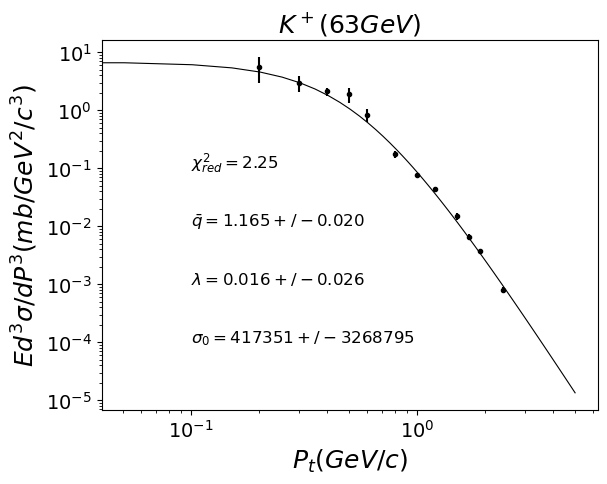} 
 \end{subfigure}	
  \centering\begin{subfigure}{}
    \includegraphics[width=0.3\linewidth]{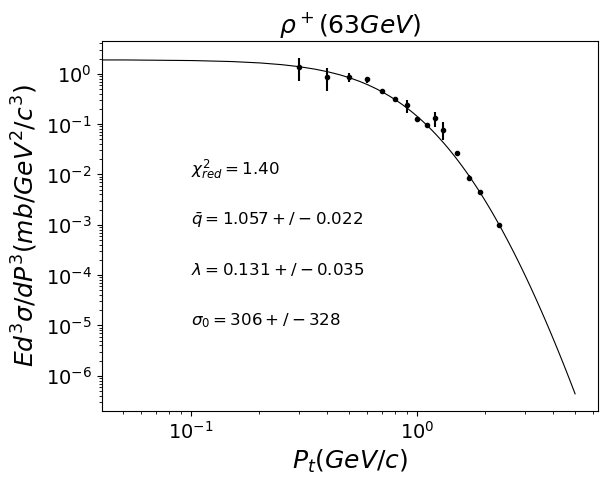} 
 \end{subfigure}	
 
 \caption{Best fit results for adjustable parameters in Scenario 1.}
	\label{fig:1-qExp}
\end{figure}

The accumulated experience of fitting $q$-exponential distributions to high-energy data in the deconfined regime has underlined the importance of addressing parameter correlations to accurately extract pertinent physical information~\cite{Sena-Deppman-2013, Marques-Andrade-Deppman-2013, Bhattacharyya_2018}. This study incorporates the methodologies established in those analyses.

\section{Results}

This section presents the results of each of the scenarios previously delineated and gives limited and preliminary discussion for each scenario. A global discussion will be presented in the next section.

\subsection{Scenario 1}

\begin{figure}[ht]
\centering
\begin{subfigure}{}
\includegraphics[width=0.4\linewidth]{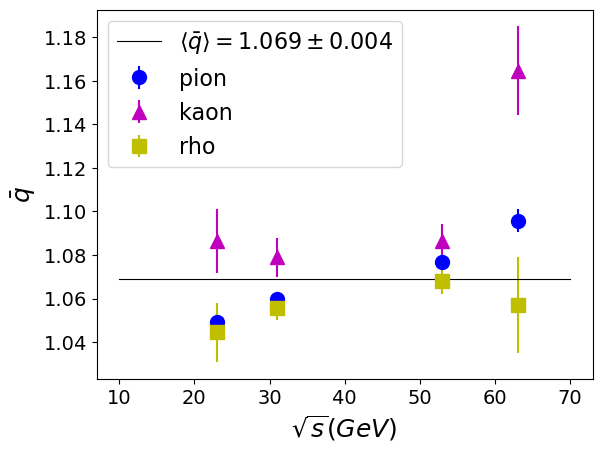} 
 \end{subfigure}
 \centering\begin{subfigure}{}
    \includegraphics[width=0.4\linewidth]{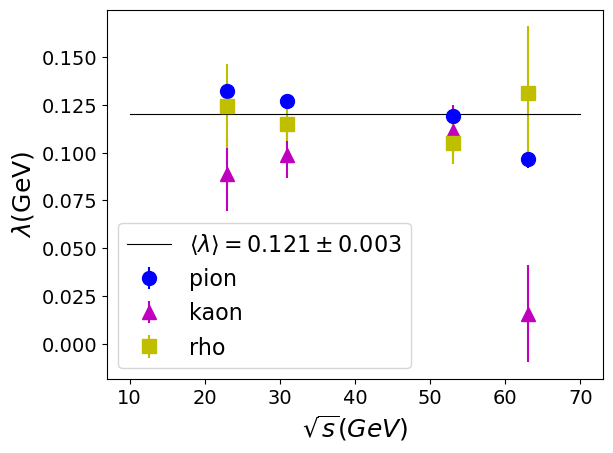}
 \end{subfigure}
 \caption{Behaviour of the parameters $q$ (left) and $\lambda$ (right) as a function of the collision energy, according to the analysis in Scenario 1.}
	\label{fig:parameters1-qExp}
\end{figure}

In the first scenario of the analysis, the best fits for the set of experimental data are displayed in Fig.~\ref{fig:1-qExp}. The single $q$-exponential function fitted the experimental data with the parameters presented in Table 1. The overall agreement is good, with reduced chi-square in the range of 1 to 4. The parameters $q$ and $\lambda$ reported in that table are plotted as a function of the collision energy in Fig.~\ref{fig:parameters1-qExp}. The worst fitting of the model to experimental data happens for $K^+$ at 63 GeV, and the same relative result will be observed in the other scenarios of the analysis. By a simple visual inspection, it is possible to observe that data set for this case departs from the general shape observed for all the other cases. This suggests that the experimental analysis and data reduction could be revisited.

The values for both $\bar{q}$ and $\lambda$ appear in the range $1.04<\bar{q}<1.10$ and $0.10<\lambda<0.15$, except for one set of data on kaon production. The mean values are $q=1.069 \pm 0.004$ and $\lambda=0.121 \pm 0.003$. 
These results show that fractal structures may be present in the hadron structure, but the values for both parameters are slightly different from those found in the QGP multiparticle production, where $q=1.14$ and $\lambda$ is close to the pion mass, $m_{\pi}=140$~MeV.
Additional information
reporting other aspects of the analysis can be found in the Supplementary Material.

The main difference between the process analyzed here and the multiparticle production at higher energies is the deconfined regime for quarks and gluons that are determinants of the minimum number of vertices in the process. While in QGP the partonic production involves a single vertex, in the hadronic collisions the confinement of quarks and gluons imposes a minimum of two vertices for any interaction to keep quarks and gluons confined. 
In the meson production studied in the present work, a minimum of three vertices are necessary, as depicted in Fig 1: one related to the meson production, where the partons produced in each of the interacting protons merge to form the new particle, and the other two at the independent interaction in each of the colliding protons, where the partons are originated.

The presence of the independent vertices makes all the differences in the process. Roughly speaking, each vertex contributes with a q-exponential, so the expected exponent in the function is twice that for the single vertex, that is, for hadronic interactions, we expect a value $\bar{q}$ such that
\begin{equation}
    \frac{1}{\bar{q}-1}=\frac{2}{q-1}\,,
\end{equation}
and using the expected value $q=1.14$ it results that $\bar{q}=1.07$, in agreement with the result of the analysis in Scenario 1 reported in Table~\ref{tab:table1}.

\begin{figure}[ht]
\centering
\begin{subfigure}{}   \includegraphics[width=0.7\linewidth]{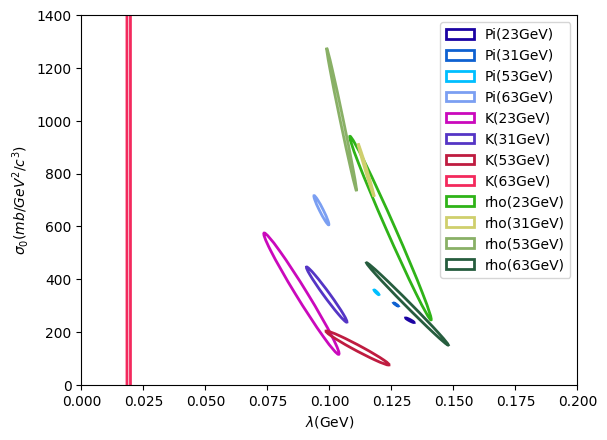}
 \end{subfigure}
  \caption{Correlations between the parameters $\sigma_o$ and $\lambda$ in Scenario 1. Ellipses with axes that are not parallel to the parameters axis unveil the correlation between the parameters.}
 \label{fig:elipseslambda-q}
 \end{figure}

The fitting of the q-exponential function faces some challenging aspects associated with the correlations in the parameter space. As an example, Fig~\ref{fig:elipseslambda-q} presents the ellipses corresponding to a constant chi-square value. Most of the ellipses have axes that are not parallel to the parameters axes, 
reflecting the existence of the correlation. For comparison, the value $q=1.07$ is indicated through a vertical line. Observe that the ellipses fall around the expected theoretical value, except in the case of the data set for $K^+$ at $63$~GeV.

The considerations made above must be clarified in many aspects, but they are indicative that the thermofractal structures manifest themselves in the confined quark matter. The main trouble with the present analysis is the use of a single form of q-exponential. If the explanation for the different value of $q$ is the presence of independent vertices, why both vertices should depend on $\varepsilon$ and not on $\varepsilon_1$ or $\varepsilon_2$. Aside from this, it is not evident that the same data that was adjusted by a single q-exponential function can be fitted as a product of q-exponential functions equally well. This will be investigated in the following scenarios.

\subsection{Scenario 2}

\begin{figure}[t]
\centering
\begin{subfigure}{}
\includegraphics[width=0.3\linewidth]{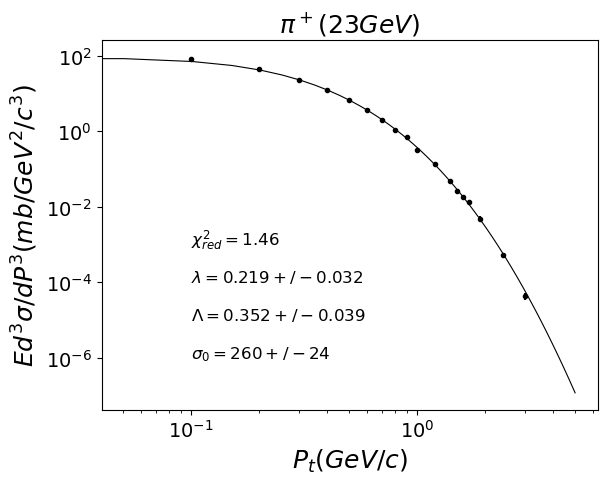} 
 \end{subfigure}
\centering
\begin{subfigure}{}
\includegraphics[width=0.3\linewidth]{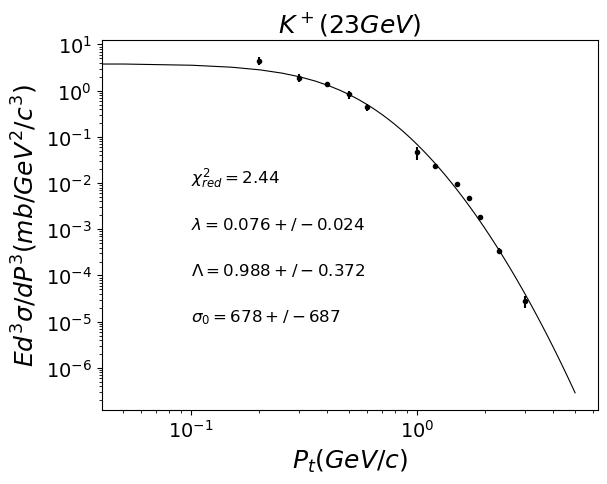} 
 \end{subfigure}
\centering
\begin{subfigure}{}
\includegraphics[width=0.3\linewidth]{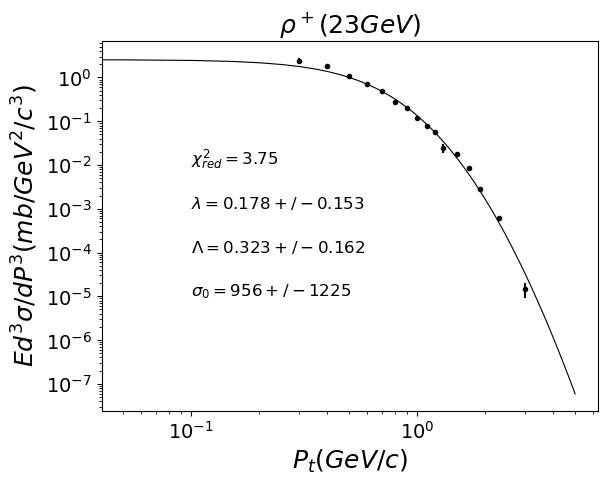} 
 \end{subfigure}
 \centering\begin{subfigure}{}
    \includegraphics[width=0.3\linewidth]{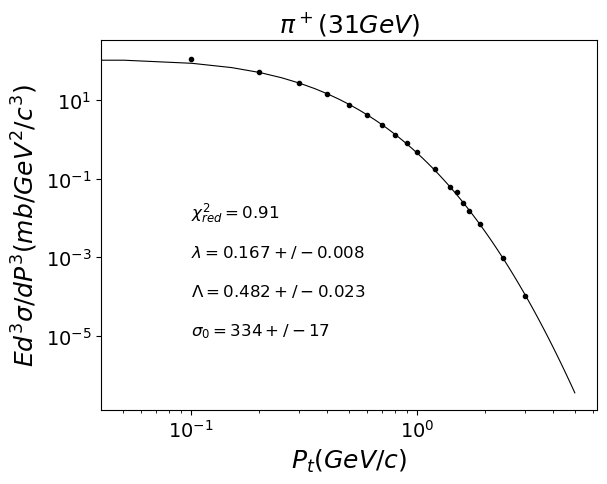}
 \end{subfigure}	
 \centering\begin{subfigure}{}
    \includegraphics[width=0.3\linewidth]{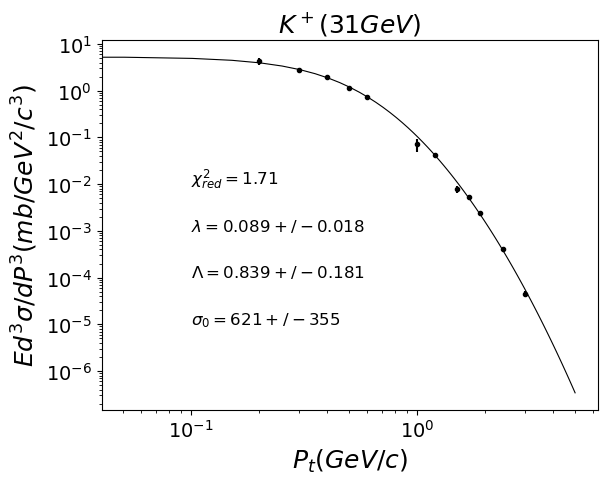}
 \end{subfigure}	
 \centering\begin{subfigure}{}
    \includegraphics[width=0.3\linewidth]{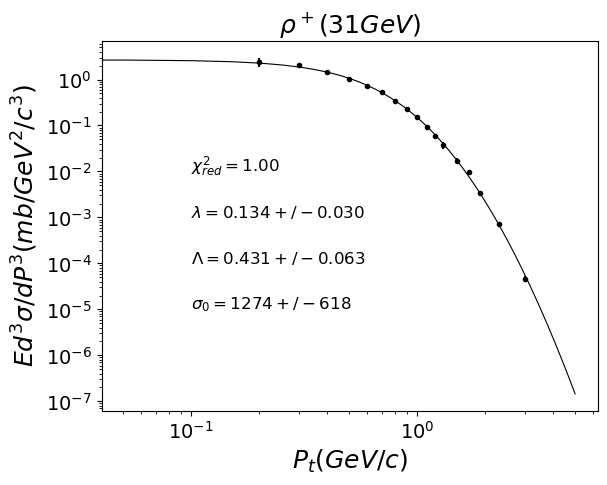}
 \end{subfigure}	
 \centering\begin{subfigure}{}
    \includegraphics[width=0.3\linewidth]{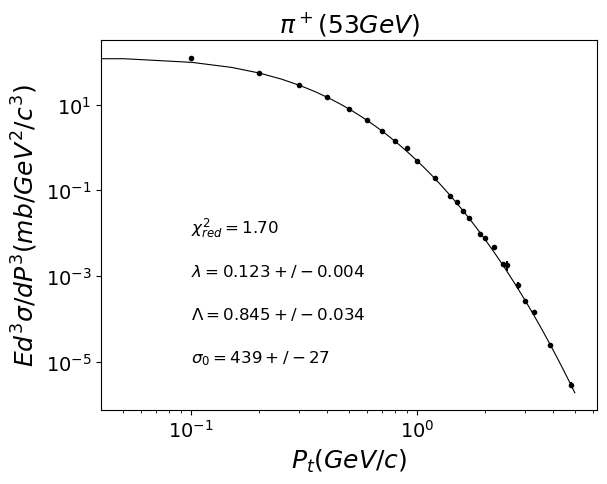} 
 \end{subfigure}	
 \centering\begin{subfigure}{}
    \includegraphics[width=0.3\linewidth]{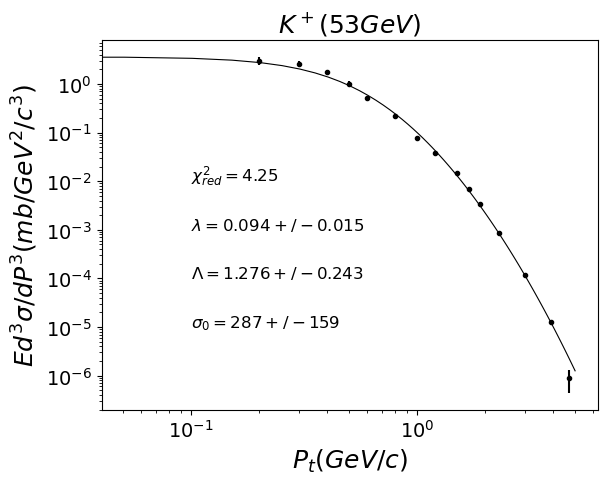} 
 \end{subfigure}	
 \centering\begin{subfigure}{}
    \includegraphics[width=0.3\linewidth]{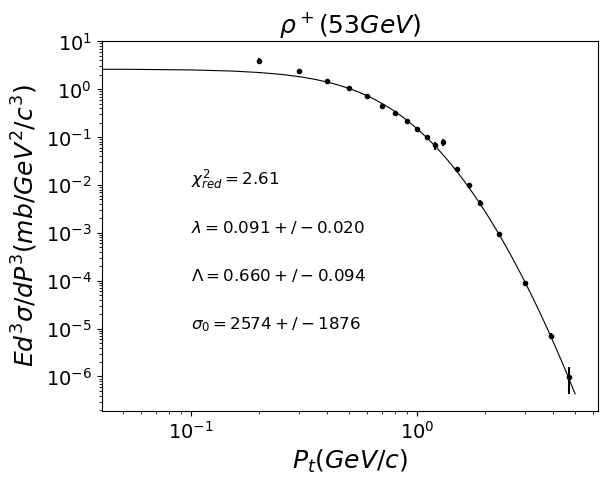} 
 \end{subfigure}	
 \centering\begin{subfigure}{}
     \includegraphics[width=0.3\linewidth]{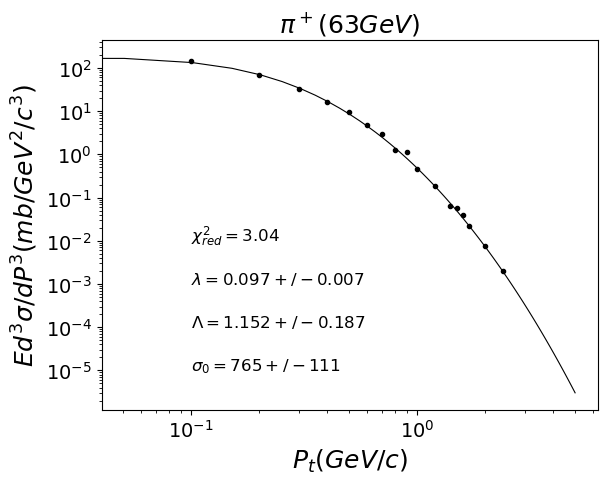} 
 \end{subfigure}	
 \centering\begin{subfigure}{}
     \includegraphics[width=0.3\linewidth]{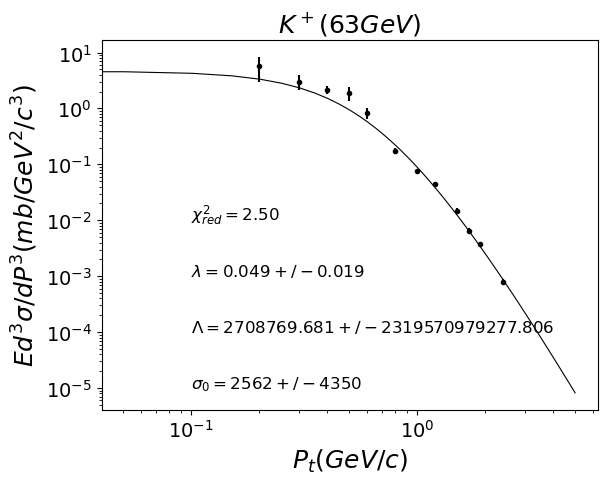} 
 \end{subfigure}	
 \centering\begin{subfigure}{}
    \includegraphics[width=0.3\linewidth]{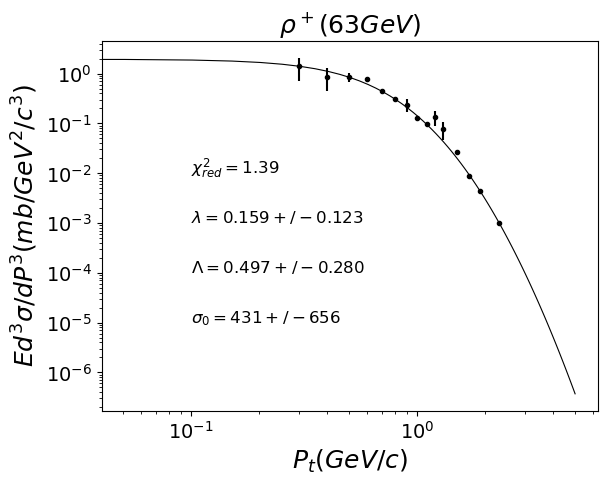} 
 \end{subfigure}	 

 \caption{Best fit results for adjustable parameters in Scenario 2.}
	\label{fig:3-qexp-free}
\end{figure}

To describe the subprocess depicted in Fig.1 using the thermofractal approach, we need to use the running coupling for each of the vertices in the process, which can be done by considering that in proton $P_1$ one pair parton-antiparton is created, contributing with a term $e_q(\varepsilon_1,\lambda_1,q) \cdot e_q(\bar{\varepsilon}_1,\lambda_1,\bar{q})$ to the cross-section, and the same happens for proton $P_2$, contributing with $e_q(\varepsilon_2,\lambda_2,q) \cdot e_q(\bar{\varepsilon}_2,\lambda_2,\bar{q})$. In the third subprocess, two partons merge to form a meson, reducing the number of q-exponentials since $e_q(\bar{\varepsilon}_1,\lambda_1,q) e_q(\bar{\varepsilon}_2,\lambda_2,q)=e_q(\varepsilon,\lambda,q)$. The parameter $\lambda$ is associated with the energy scale involved in the meson production, while the parameter $\Lambda$ is associated with the colliding hadrons energy scale. Due to the energy-momentum conservation, the constraint $\bar{\varepsilon}_{1(2)}=\varepsilon_{1(2)}$ must be satisfied. The use of the index $\bar{q}$ for the vertices in the colliding protons is due to the fact that the vertex occur in the complex environment of the hadron structure, which is supposed to be in equilibrium~\cite{Megias2015}.

With the considerations made above, the differential cross-section will be proportional to the product of three q-exponential functions, i.e.,
\begin{equation}
    \frac{d^3 \sigma}{dp^3}= \sigma_o e_q(\varepsilon_1,\lambda_1,\bar{\bar{q}}) e_q(\varepsilon_2,\lambda_2,\bar{\bar{q}}) e_q(\varepsilon,\lambda,q) \,. \label{3qexp}
\end{equation}
The parameters $\lambda_1$ and $\lambda_2$ are related to the internal characteristics of the colliding protons. Due to the symmetries of the two protons involved in the process, $\lambda_1=\lambda_2=\Lambda$, and the expression above reduces to the one proposed for Scenario 2 of the analysis, given in Eq.~\ref{eq:mod2}

The same sets of data were adjusted by the cross-section in Eq.~\ref{3qexp} with $q=1.14$ and the parameters $\sigma_0$, $\Lambda$ and $\lambda$ adjustable. The results of the analysis for Scenario 2 are displayed in Fig.~\ref{fig:3-qexp-free}. The formula can fit well the data, resulting in a fairly low chi-square for most of the cases, as reported in Table~\ref{tab:table2}. This is itself a promising result since the physical role of each term in the cross-section can be clearly understood. Reinforcing the physical significance of the results, the values for $\Lambda$ and $\lambda$ can be compared to those found in the higher energy processes. The best-fit values for the scale parameters are displayed in Fig.~\ref{fig:lambdas-free}.

For $\lambda$ and $\Lambda$ there is no evident dependence on the produced meson mass. The values for $\lambda$ fall in the range $0.5~GeV$ and $0.25~GeV$, while for $\Lambda$ the values vary on a broader range, from $0.2~GeV$ to $1.4~GeV$. Both parameters present some dependence on the collision energy, which results to be stronger for $\Lambda$ than for $\lambda$. 

\begin{figure}[t]
\centering
\begin{subfigure}{}
     \includegraphics[width=0.6\linewidth]{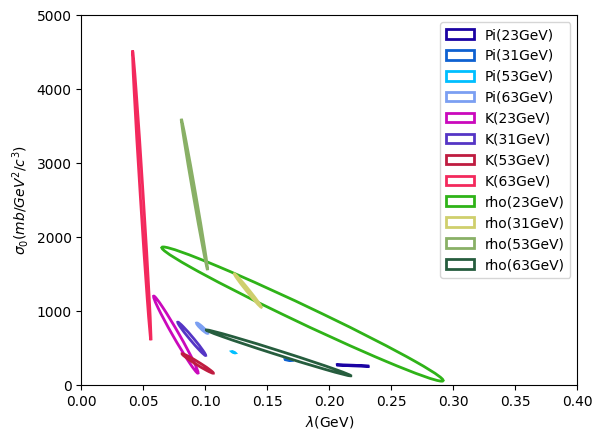}   
\end{subfigure}
\centering
\begin{subfigure}{}
     \includegraphics[width=0.6\linewidth]{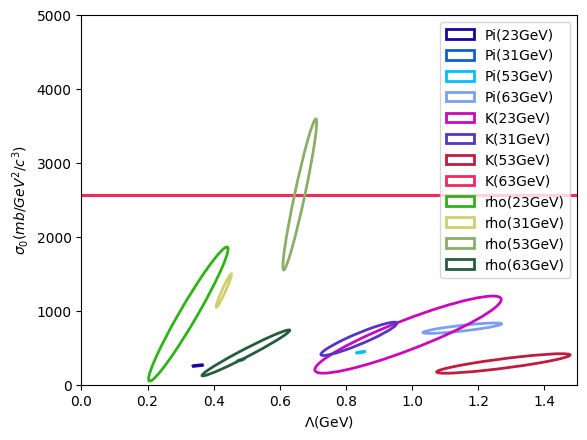}      
\end{subfigure}

 \caption{Visual representation of the correlations in the parameter space during the fitting process. The plot shows the correlations for $\sigma_0$ with $\lambda$ and $\Lambda$, as used in Scenario 2 of the analysis.}
\label{fig:ellipse-lambdaLambda}
\end{figure}

It is important to notice that, as happens in high-energy studies, the fitting process for the hadron regime analysis performed here needs to address the difficulty of correlation among the adjustable parameters~\cite{Sena-Deppman-2013,Marques-Andrade-Deppman-2013,Bhattacharyya_2018}. The correlations are illustrated in the $\lambda$ vs $\Lambda$ projection of the parameter space shown in Fig.~\ref{fig:ellipse-lambdaLambda}, where the typical ellipses already observed in analyses of multiparticle production can be observed. The average value $\langle \lambda \rangle = 0.122 \pm 0.007$ is a reasonable value for almost all sets analyzed and is close to the average values obtained in Scenario 1, whilst for $\Lambda$ the values are spread over a larger range around the average $\langle \Lambda \rangle=0.55 \pm 0.06 ~GeV$. Similar correlations exist between other pairs of parameters, leading to a small variation of the chi-square for the best fit when some of the parameters are fixed at reasonable values, as observed in Tables~\ref{tab:table1} to~\ref{tab:table3}. Additional information on the correlations in the parameter space can be found in the Supplementary Material. Therefore, to simplify the analysis, it is reasonable to assume that $\lambda$ is constant. This modification corresponds to Scenario 3 of the analysis.

The average $\langle \lambda \rangle$ is close to the pion mass, so the parameter $\lambda$ will be fixed to the pion mass, $m_{\pi}=0.140~GeV$. Assuming this value, this work departs from a purely statistical analysis, but gains in the physical interpretation of the result.

At this point, it is important to stress that other possibilities of analysis are present and will be left for future work.

\begin{figure}[t]
\centering
\begin{subfigure}{}
   \includegraphics[width=0.4\linewidth]{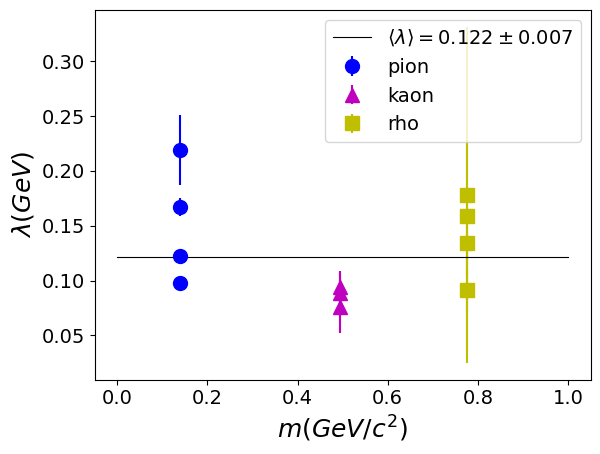} 
 \end{subfigure}	
 \centering\begin{subfigure}{}
     \includegraphics[width=0.4\linewidth]{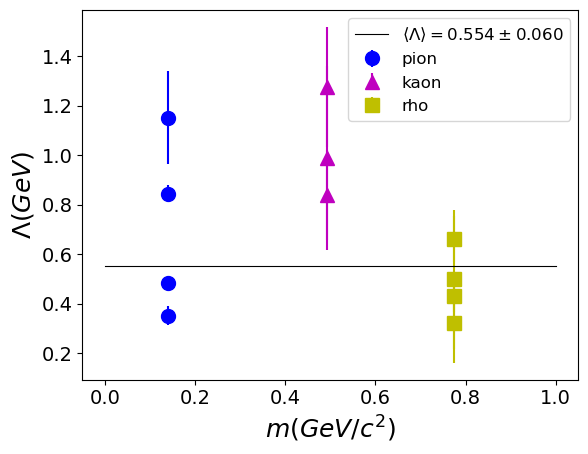} 
 \end{subfigure}	
 \caption{Behaviour of the parameters $\lambda$ (left) and $\Lambda$ (right) as a function of produced meson mass, according to the analysis in Scenario 2.}
	\label{fig:lambdas-free}
\end{figure}

\subsection{Scenario 3}

\begin{figure}[t]
\centering
\begin{subfigure}{}
\includegraphics[width=0.3\linewidth]{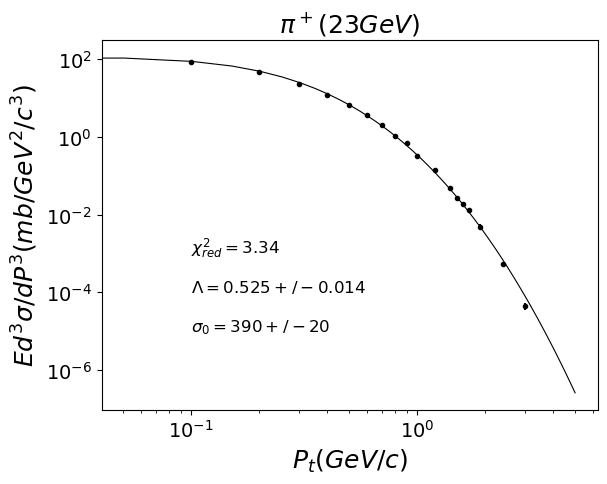} 
 \end{subfigure}
\centering
\begin{subfigure}{}
\includegraphics[width=0.3\linewidth]{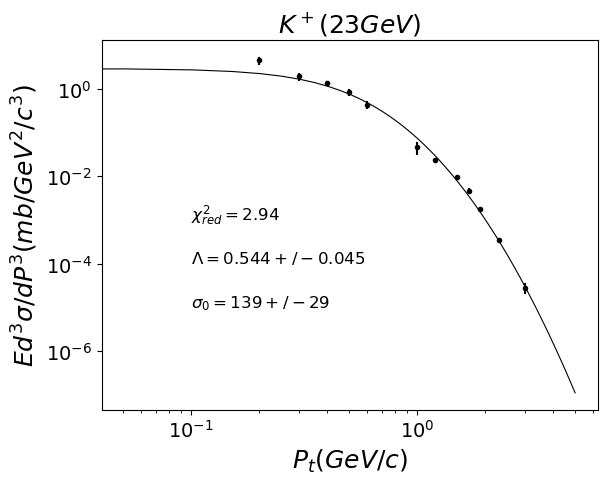} 
 \end{subfigure}
\centering
\begin{subfigure}{}
\includegraphics[width=0.3\linewidth]{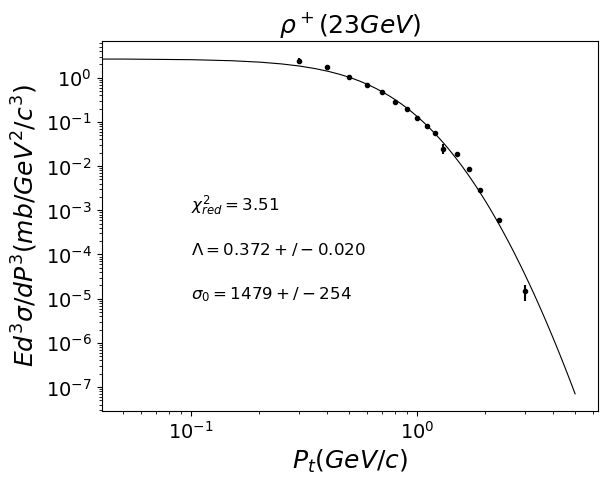} 
 \end{subfigure}
 \centering\begin{subfigure}{}
    \includegraphics[width=0.3\linewidth]{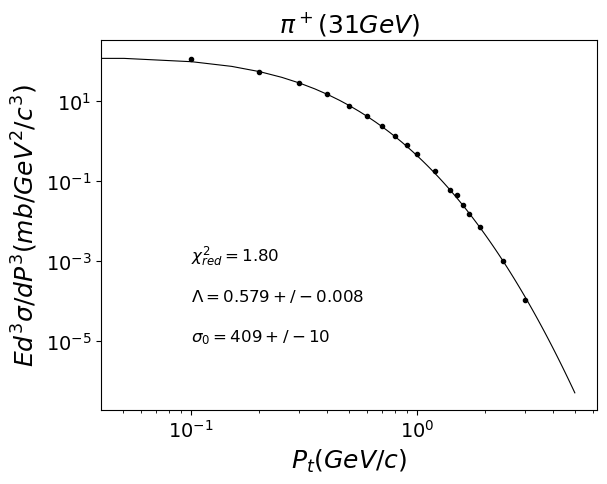}
 \end{subfigure}	
 \centering\begin{subfigure}{}
    \includegraphics[width=0.3\linewidth]{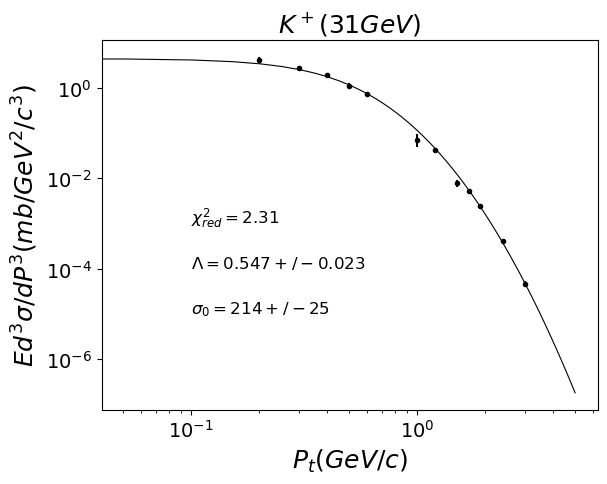}
 \end{subfigure}
 \centering\begin{subfigure}{}
    \includegraphics[width=0.3\linewidth]{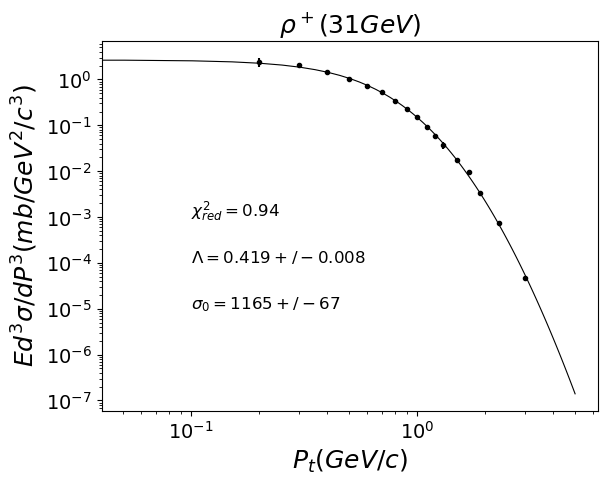}
 \end{subfigure}	
 \centering\begin{subfigure}{}
    \includegraphics[width=0.3\linewidth]{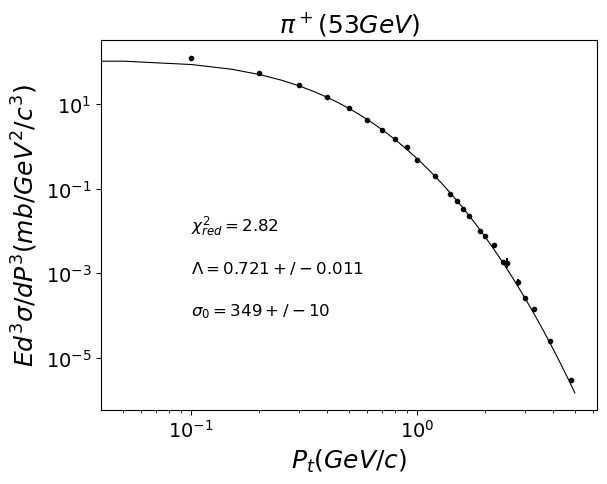} 
 \end{subfigure}	
 \centering\begin{subfigure}{}
    \includegraphics[width=0.3\linewidth]{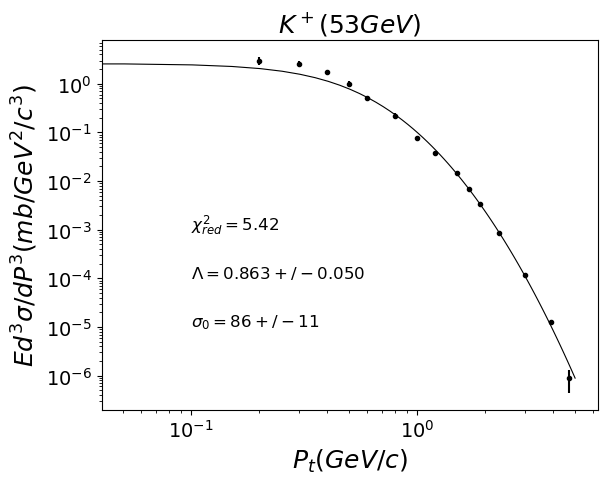} 
 \end{subfigure}
 \centering\begin{subfigure}{}
    \includegraphics[width=0.3\linewidth]{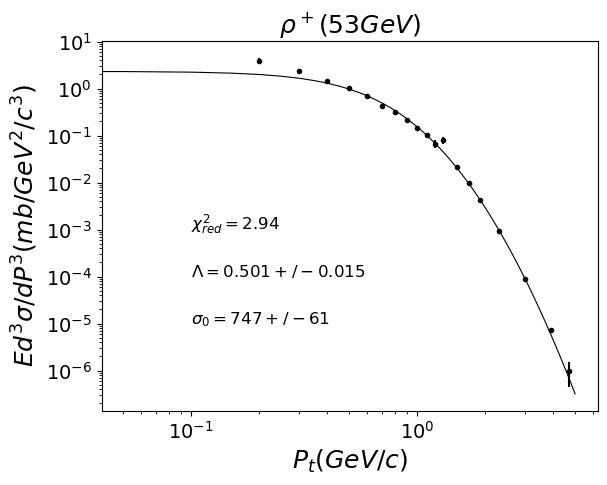} 
 \end{subfigure}	
 \centering\begin{subfigure}{}
     \includegraphics[width=0.3\linewidth]{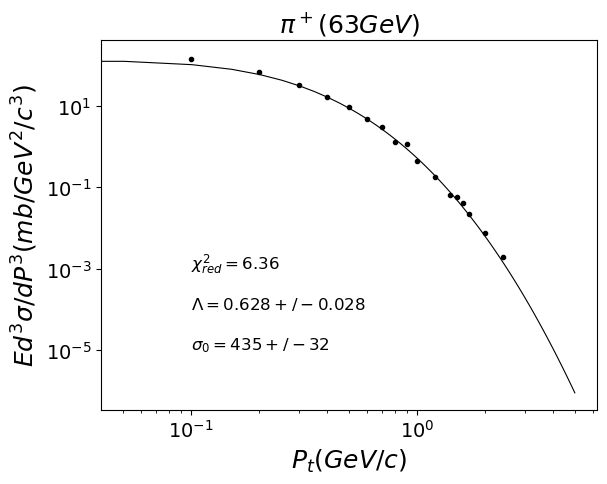} 
 \end{subfigure}	
 \centering\begin{subfigure}{}
     \includegraphics[width=0.3\linewidth]{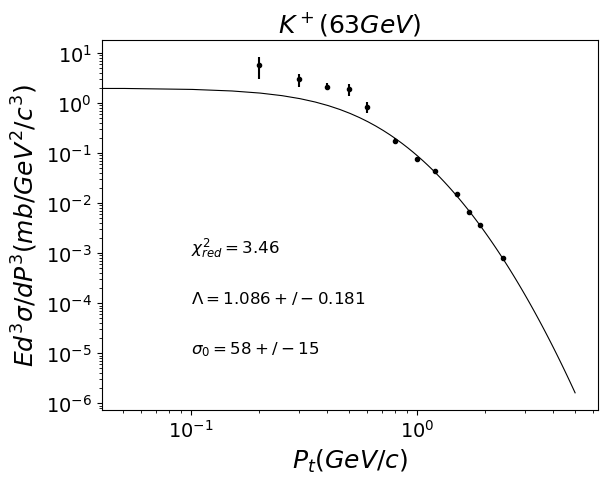} 
 \end{subfigure}
 \centering\begin{subfigure}{}
    \includegraphics[width=0.3\linewidth]{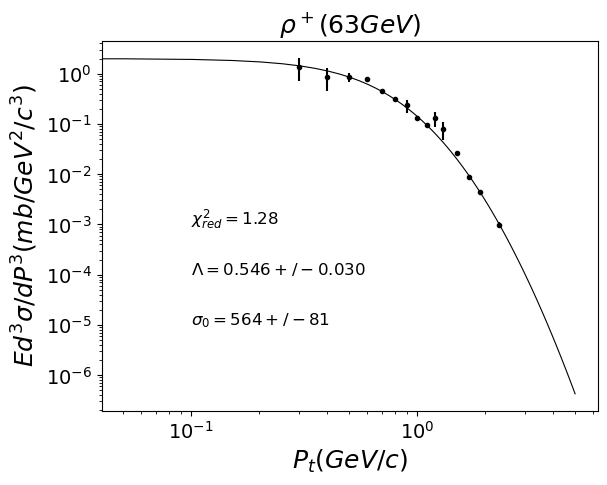} 
 \end{subfigure}	 

 \caption{Best fit results for adjustable parameters in Scenario 3.}
	\label{fig:3-qexp-fixed}
\end{figure}

\begin{figure}[t]
\centering
 \centering\begin{subfigure}{}
     \includegraphics[width=0.5\linewidth]{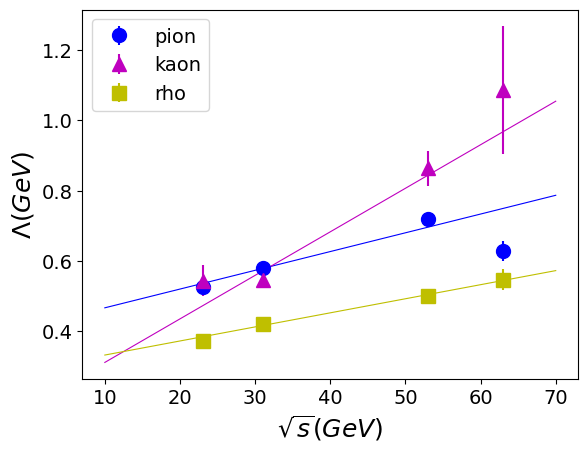} 
 \end{subfigure}	
 \caption{Behaviour of the parameter 
 $\Lambda$ as a function of
 collision energy for the different meson species according to the analysis in Scenario 3. Straight lines are fitted to the data for guiding the eyes.}
	\label{fig:lambda-fixed}
\end{figure}

The third and last scenario of the analysis represents the most reliable description of the process presented in this work. While the number of adjustable parameters was three in the two previous scenarios, the third scenario has only two free parameters in the fitting. The other parameters were fixed to values corresponding to well-known physical quantities, such as the pion mass for $\lambda$ and the thermofractal QCD value for $q$, which has been found in agreement with the value resulting from the experimental analysis. With fewer adjustable parameters, the model is subjected to a stronger test when compared with the experimental data, and the best-fit values for the parameter tend to present more significant physical content.

The model can describe accurately all sets of experimental data, as shown in the plots in Fig.~\ref{fig:3-qexp-fixed}. The best-fit values for $\Lambda$ are given in Table~\ref{tab:table3} and in the plots presented in Fig.~\ref{fig:lambda-fixed}. The parameter $\Lambda$ shows dependence on the produced meson and on the collision energy. In the range analysed the behaviour of the scale $\Lambda$ is approximately linear for all mesons studied.

\begin{figure}[ht]
\centering
\begin{subfigure}{}
\includegraphics[width=0.7\linewidth]{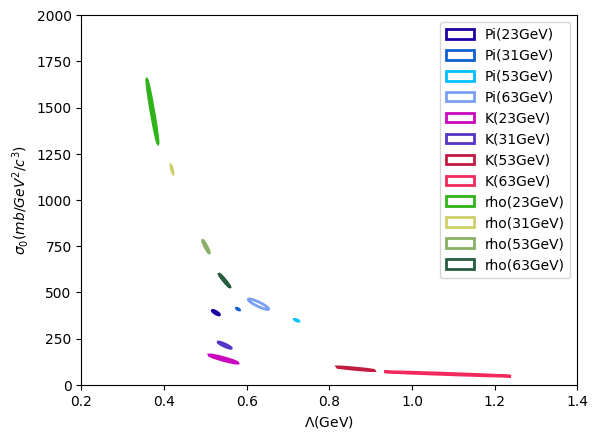} 
 \end{subfigure}
 \caption{Correlations between the parameters $\sigma_o$ with $\Lambda$ in Scenario 3.}
	\label{fig:SM1} \label{fig:ellipsesSigmaLambda-S3}
\end{figure}

The study of correlations in the parameter spaces brings evidence of strong correlations involving the adjustable parameters, which can cause some difficulties in obtaining the real values for the adopted parameters in each scenario of the analysis. In particular, Fig.~\ref{fig:ellipsesSigmaLambda-S3} shows that the correlations between $\sigma_o$ and $\Lambda$, the only free parameters in this scenario of the analysis, roughly distribute along a power-law curve. This form is due to the known relationship between the multiplicative factor of the q-exponential function and the multiplicative factor in the argument of the function, as demonstrated in Ref.~\cite{TsallisBookV2} of the work, on page 35.

\section{Discussion and conclusion}

The results obtained here provide at least some preliminary answers to the questions posed in the Introduction. The findings show that fractal structures may be present in the hadron structure, as the momentum distributions of the produced mesons in $pp$ collisions can be accurately described by q-exponential functions. The analysis progressively unveils the role of the fractal structure in the subprocesses involved in the hadronic interactions.

In the first scenario of the analysis, a naive model of the fractal structure was applied, and it was already possible to observe a good fit of the data with values for the parameters $q$ and $\lambda$ that seemed to be consistent with those observed in high-energy multiparticle production. Even with such a naive approach, it was possible to find evidence of fractal structures with characteristics similar to those found in the deconfined regime. Interpreting the results from Scenario 1, it became evident that at least two vertices must be involved in the confined regime to ensure the confinement of quarks and gluons.

In Scenario 2, the subprocess involving partons from both colliding protons was described more accurately than in the first scenario, introducing a new parameter, $\Lambda$, while fixing $q=1.14$. The parameter $\Lambda$ represents the energy scale of the colliding protons, while the parameter $\lambda$ represents the energy scale of the produced meson. The results obtained in this scenario showed that the model for the momentum distribution can describe the data, and the parameter $\lambda$ could be considered independent of the collision energy and the produced meson species, although a weak dependence on those variables cannot be completely ruled out. On the other hand, variations in the parameter $\Lambda$ with energy and particle mass were more evident.

In Scenario 3, a more realistic model was adopted, with only two adjustable parameters, $\sigma_o$ and $\Lambda$, while keeping $q$ fixed to the same value as before, i.e. q =1.14, and fixing $\lambda$ to the pion mass. Even with these constraints, it was possible to describe the data, and the behaviour of $\Lambda$ with the particle species and the collision energy was analyzed.

The main aspects that can be inferred from the more realistic scenario are as follows: the scale $\Lambda$ depends on the particle species and the collision energy. While the energy dependence, for the range considered in the present work, can be considered approximately linear for all particles analyzed, the rate of increase of $\Lambda$ with the collision energy depends on the quark composition of the created meson. Specifically, for pion and rho, which are formed by $(u,\bar{d})$ valence quarks, $\Lambda$ increases at approximately the same rate, whereas for the Kaon, formed by $(u,\bar{s})$ valence quarks, the parameter increases at a faster rate. The different rates of increase in $\Lambda$ cannot be attributed solely to the differences in meson masses, since pion and rho have similar slopes with different masses, thereby it may be associated with the different quantum numbers of the mesons.

\begin{figure}[t]
\centering
 \centering\begin{subfigure}{}
     \includegraphics[width=0.4\linewidth]{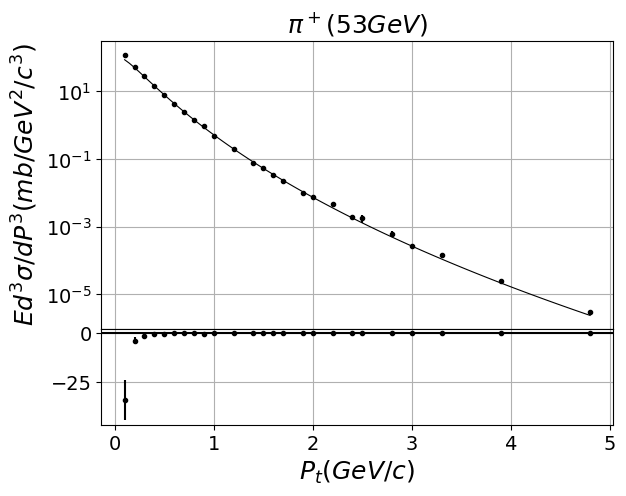} 
 \end{subfigure}	
\centering\begin{subfigure}{}
     \includegraphics[width=0.4\linewidth]{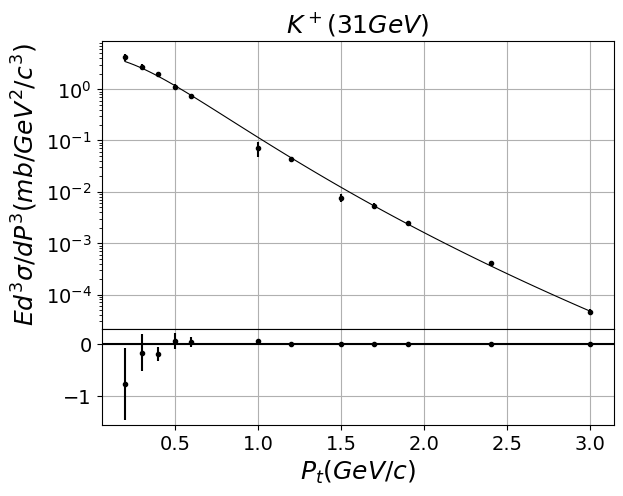} 
 \end{subfigure}	
\centering\begin{subfigure}{}
     \includegraphics[width=0.4\linewidth]{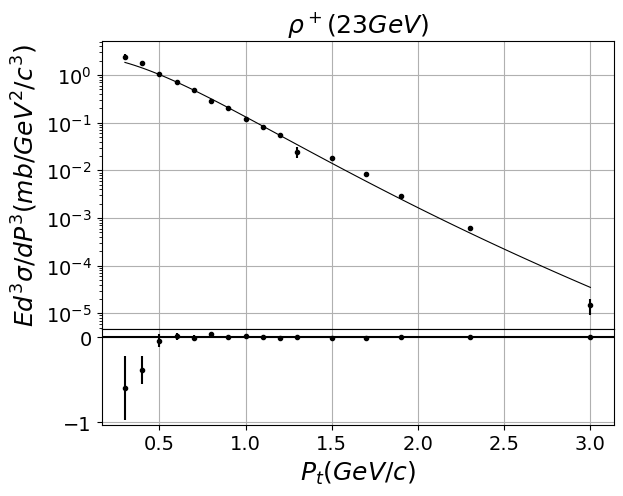} 
 \end{subfigure}	
 \caption{Analysis of the residues in the fitting process for typical cases using the Scenario 3 analysis. Additional results for all scenarios can be found in the Supplementary Material.}
	\label{fig:residues}
\end{figure}

Being more realistic, the third scenario allows the investigation, in the hadron structure, of one surprising, and yet not understood, effect that is observed in high-energy collisions is the presence of log-periodic oscillations in the transverse momentum distributions. The oscillations appear in the normalized residues after the fittings of the experimental data with the S$_q$ distributions. This work investigates the presence of the same kind of oscillations in the hadronic case. Fig.~\ref{fig:residues} reports typical results of the analysis, and the complete report of the residues can be found in the Supplementary Material. Although the shorter range of momentum distribution limits the analysis, the results obtained do not allow one to conclude that the same oscillations appear in the confined regime. The absence of the log-periodic oscillations in the hadronic regime can be indicative of the connection between the oscillation and collective behaviour of the QGP, possibly associated with the flow.

The conclusions one can infer from the study presented in this work are the following: It is likely that the fractal structure is present in confined hadronic matter with characteristics similar to those observed in higher energy collisions, where the deconfinement of quarks and gluons is expected. The significance of this finding lies in demonstrating the persistence of the fractal structure through the confined/deconfined phase transition.

The energy scale associated with the produced meson is the pion mass, and it is similar QGP freeze-out temperature. This is another evidence that the same fractal structure is present in hadrons as well as in QGP.

The scale $\Lambda$ depends on the meson mass, on the meson quark content and on the collision energy. The rate of the scale increase with energy seems to be independent of the quark content, at least considering the limited experimental data analyzed here.

This work presents a systematic analysis of meson production in proton-proton ($pp$) collisions within a specific energy range. The main objective is to demonstrate that the thermofractal structures observed in multiparticle production at higher energies are also present in the confined region. By investigating fractal structures in hadrons, the research contributes significantly to understanding the presence of fractal effects in the hadronic processes, illuminating the physical meaning of the fractal parameters. It is observed that certain fractal parameters, specifically $q$ and $\lambda$, are universal and independent of whether the hadronic matter is in the confined or deconfined regime, while $\sigma_o$ and $\Lambda$ may be influenced by the quark structure of the produced meson.

The study opens up opportunities for further investigation in terms of collision energy and particle species. Fractal characteristics, such as scale invariance and self-similarity, may have significant effects on observables like Parton Distribution Function (PDF)~\cite{Dulat2016-po}, Transverse Momentum Distribution (TMD)~\cite{Echevarria2016-wo,Anselmino2014-iv}, Generalized Momentum Distribution (GPD)~\cite{Belitsky2005-sq,Diehl2016-in} and other distributions that can be measured experimentally. These effects will be explored more thoroughly in future experiments using the Electron-Ion Collider (EIC)~\cite{Frankfurt2005-ee,Chen2020-bm}. In Astrophysics, the hadronic composition and processes are relevant in the study of massive objects~\cite{Wang2023,RAZEIRA2007,Gervino2013} and can benefit from the present results. 

The z-scaling is a phenomenological approach to meson production that evidences the scaling properties of the process\cite{Z-Scaling,Z-Scaling-in-pA,Generalized-Z-Scaling1}. The comparison between the thermofractal approach and the z-scaling can further clarify the underlying physics of the hadronic collisions.

Interestingly, the fact that the phase transition between confined and deconfined regimes does not disrupt the fractal structure may offer constraints on the nature of the phase transition itself.

In conclusion, the results presented in this work lay the groundwork for future research aimed at understanding the intricate fractal nature of hadrons and its implications for various physical distributions critical in high-energy physics. The findings may lead to advancements in our comprehension of the underlying structure of hadrons and aid in the interpretation of experimental data from high-energy collisions.

\section{Acknowledgements}
The authors thank Prof. Otaviano A. M. Helene for an insightful discussion on the correlations among parameters in the fitting process. 
A.D. is supported by the Project INCT-FNA (Instituto Nacional de Ci\^encia e Tecnologia - F\'{\i}sica Nuclear Aplicada) Proc. No. 464898/2014-5, and by the Conselho Nacional de Desenvolvimento Cient\'{\i}fico e Tecnol\'ogico (CNPq-Brazil), grant 306093/2022-7.  C.T. is partially supported by CNPq and Faperj (Brazilian agencies).

\begin{table}[p]
    \centering
    \begin{tabular}{|c | c | c | c | c|} 
     \hline
     \multicolumn{5}{|c|}{$\pi^+$} \\
     \hline
     $\sqrt{s}(GeV)$ & $\sigma_0(mb/GeV^2/c^3)$ & $\lambda(GeV)$ &  $\bar{q}$ & $\chi^2_{red}$ \\ [0.5ex] 
     \hline
     23 & 244(20) & 0.132(4) & 1.049(4) & 1.59 \\ 

     31 & 304(15) & 0.127(2) & 1.060(2) & 1.21 \\ 

     53 & 350(21) & 0.119(2) & 1.077(2) & 2.25\\ 
  
     63 & 661(90) & 0.097(5) & 1.096(5) & 2.87\\ 
     \hline
     \multicolumn{5}{|c|}{$K^+$} \\
     \hline
     23 & 345(288) & 0.09(2) & 1.09(2) & 2.68 \\ 

     31 & 341(155) & 0.10(1) & 1.079(9) & 1.93 \\ 

     53 & 139(69) & 0.11(1) & 1.086(8) & 5.33\\ 
  
     63 & 0.4(3)E6 & 0.02(3) & 1.17(2) & 2.25\\ 
     \hline
     \multicolumn{5}{|c|}{$\rho^+$} \\
     \hline
     23 & 593(466) & 0.13(2) & 1.04(1) & 3.81 \\ 

     31 & 813(263) & 0.115(9) & 1.056(5) & 1.09 \\ 

     53 & 1005(479) & 0.11(1) & 1.068(6) & 2.82\\ 
  
     63 & 306(328) & 0.13(4) & 1.06(2) & 1.40\\ 
    \hline
    \end{tabular}
    \caption{Best-fit results for adjustable parameters in Scenario 1.}
    \label{tab:table1}
\end{table}

\begin{table}[p]
    \centering
    \begin{tabular}{|c | c | c | c | c|} 
     \hline
     \multicolumn{5}{|c|}{$\pi^+$} \\
     \hline
     $\sqrt{s}(GeV)$ & $\sigma_0(mb/GeV^2/c^3)$ & $\lambda(GeV)$ & $\Lambda(GeV)$ & $\chi^2_{red}$ \\ [0.5ex] 
     \hline
     23 & 260(24) & 0.22(3) & 0.35(4) & 1.46 \\ 

     31 & 334(17) & 0.167(8) & 0.48(2) & 0.91 \\ 

     53 & 439(27) & 0.123(4) & 0.85(3) & 1.70\\ 
  
     63 & 765(111) & 0.097(7) & 1.2(2) & 2.87\\ 
     \hline
     \multicolumn{5}{|c|}{$K^+$} \\
     \hline
     23 & 678(687) & 0.08(2) & 1.0(4) & 2.44 \\ 

     31 & 621(355) & 0.09(2) & 0.8(2) & 1.71 \\ 

     53 & 287(59) & 0.09(2) & 1.3(2) & 4.25\\ 
  
     63 & 3(4)E3 & 0.05(2) & 0(2e12) & 2.50\\ 
     \hline
     \multicolumn{5}{|c|}{$\rho^+$} \\
     \hline
     23 & 956(1225) & 0.2(2) & 0.3(2) & 3.75 \\ 

     31 & 1274(618) & 0.13(3) & 0.43(6) & 1.00 \\ 

     53 & 2574(1876) & 0.09(2) & 0.66(9) & 2.61\\ 
  
     63 & 431(656) & 0.2(1) & 0.5(3) & 1.39\\ 
    \hline
    \end{tabular}
    \caption{Best-fit results for adjustable parameters in Scenario 2.}
    \label{tab:table2}
\end{table}

\begin{table}[h!]
    \centering
    \begin{tabular}{|c | c | c | c |} 
     \hline
     \multicolumn{4}{|c|}{$\pi^+$} \\
     \hline
     $\sqrt{s}(GeV)$ & $\sigma_0(mb/GeV^2/c^3)$ & $\Lambda(GeV)$ & $\chi^2_{red}$ \\ [0.5ex] 
     \hline
     23 & 390(20) & 0.53(1) & 3.34 \\ 

     31 & 409(10) & 0.579(8) & 1.80 \\ 

     53 & 349(10) & 0.72(1) & 2.82 \\ 
  
     63 & 435(32) & 0.63(3) & 6.36 \\ 
     \hline
     \multicolumn{4}{|c|}{$K^+$} \\
     \hline
     23 & 139(29) & 0.54(4) & 2.94  \\ 

     31 & 214(25) & 0.55(2) & 2.31  \\ 

     53 & 86(11) & 0.86(5) & 5.42 \\ 
  
     63 & 58(15) & 1.1(2) & 3.46 \\ 
     \hline
     \multicolumn{4}{|c|}{$\rho^+$} \\
     \hline
     23 & 1479(254) & 0.37(2) & 3.51  \\ 

     31 & 1165(67) & 0.419(8) & 0.94  \\ 

     53 & 747(61) & 0.50(2) & 2.94 \\ 
  
     63 & 564(81) & 0.55(3) & 1.28 \\ 
    \hline
    \end{tabular}
    \caption{Best-fit results for adjustable parameters in Scenario 3.}
    \label{tab:table3}
\end{table}

\bibliographystyle{ieeetr}
\bibliography{FractalHadron.bib}

 \end{document}